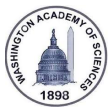

# PROPOSED MIAMI IMPACT CRATER IDENTIFIED AS A SOLUTIONAL DOLINE OF OOLITIC LIMESTONE

ANTONIO J. PARIS, RYAN ROBERTSON, & SKYE SCHWARTZ

PLANETARY SCIENCES, INC.

## ABSTRACT

This investigation addresses the discovery of a proposed impact crater located off the coast of Miami, FL under the North Atlantic Ocean. A preliminary analysis of bathymetry data obtained from the National Oceanic and Atmospheric Administration (NOAA) implied a morphology consistent with a complex crater produced by a hypervelocity impact event of extraterrestrial origin. The proposed impact crater's features include a central peak, concentric rings, and an ejecta field to the northwest. Analysis of geological data from the US Geological Survey (USGS) places the strata overlying the proposed impact site as Miami Limestone (Pleistocene), accumulated during Marine Isotope Stage 5e, thereby placing the maximum age of the proposed impact crater at ~80 ka to ~130 ka. Three other competing hypotheses for the formation of the structure, namely a controlled maritime explosion, radial lava flow from volcano, or a depressed bioherm, doline, or karst (i.e., solutional depression) were explored throughout the investigation. To confirm the proposed structure as an impact crater, an *in-situ* underwater expedition was organized by Planetary Sciences, Inc. specifically to ascertain whether planar formations, shatter cones, and shock metamorphic and/or other meteoritic properties were present. After analyzing the geological samples collected at the proposed impact crater, examining the morphology of analogous geologic structures, and evaluating competing hypotheses, we conclude that the structure is a solutional doline formed by the uneven dissolution of the Miami Limestone, and, accordingly, do not recommend that the structure be indexed in the Earth Impact Database.

## INTRODUCTION

This investigation is focused on an impact crater initially proposed by Cory Boehne in 2012. The impact crater is located 8 m under the North Atlantic Ocean at 25° 44' 59.31" N and 80° 7' 21.78" W. It is 1.20 km southeast of the entrance of Government Cut—a manmade shipping channel between Miami Beach and Fisher Island, FL (Figure 1). Although an initial inquiry was completed by Charles O'Dale in 2012,[1] the identification of the structure remained unresolved and, as a result, it was not indexed in the official Earth Impact Database.[2]

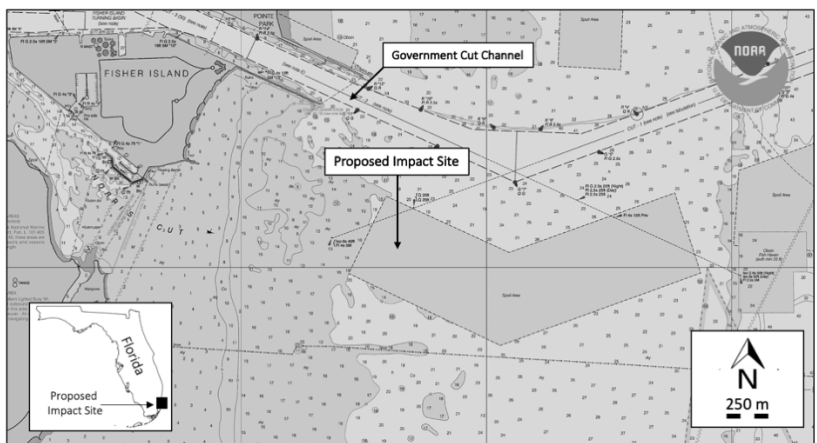

Figure 1: Location of the proposed impact crater (Source: NOAA Chart 11468)

## DATA COLLECTION

The geological data used throughout this investigation was obtained from the USGS, ArcGIS and the Association of American State Geologists (AASG).[3] Indices checks in the National Geologic Map Database (NGMDB) provided supplementary data, such as a remote sensing inventory of the proposed impact site and scholarly sources of information focused on the geologic history of southern Florida. The geologic maps and data contributed to the NGMDB have been standardized in accordance with the Geologic Mapping Act of 1992, section 31f(b), and they are widely accepted standards.

The bathymetry data used to analyze the proposed impact site was available through the National Oceanic and Atmospheric Administration (NOAA), which collects and archives multibeam and hydrographic lidar data from the earliest commercial installations. Indices checks for metadata through the NOAA's Bathymetric Data Viewer (BDV) provided a Shallow Water Multibeam Hydrographic and Side Scan Sonar Survey (Registry No. H11898) of the proposed impact site.[4] The purpose of the survey was to provide the NOAA with modern, accurate hydrographic survey data to update the nautical charts of the North Atlantic Ocean east of Key Biscayne, FL.[5] Two hundred percent side scan sonar (SSS) coverage, along with concurrent shallow water multibeam echo sounder (SWMB) coverage were acquired with set line spacing to water depths of 20 m or shallower.[6] According to the NOAA, all equipment was installed, calibrated, and operated in accordance with the requirements set forth in its Data Acquisition and Processing Reports procedures.

To complement this investigation, an underwater expedition comprised of a SCUBA diving team was conducted *in situ*. The dive plan involved taking measurements of the prominent features of the proposed impact crater (e.g., the central peak, concentric rings, and the ejecta field), underwater photography, and the collection of geological samples for planar formation, shatter cone, and shock metamorphic analysis. These features are uniquely characteristic of the intense shock of a large meteorite impact. Volcanic explosions do not generate such shocks and these features. Aerial imagery, moreover, was acquired through the use of a crewless aerial vehicle (UAV) operated over the proposed impact site. The UAV offered a powerful camera on a 3-axis stabilized gimbal that recorded video at 4k resolution up to 60 frames per second and featured real-glass optics that captured aerial imagery at 12 megapixels from an altitude of up to 800 m and a range of up to 7 km.[7]

## GEOLOGY OF THE IMPACT SITE

Geologically, the overlying strata at the impact site is Miami Limestone (formally known as Miami Oölite), and it covers a large portion of the southern tip of Florida, at or near the surface, along the Atlantic Coastal Ridge (Figure 2).[8] The formation was deposited during the Sangamonian interglacial and Wisconsin glacial stages, when the proposed impact site was under a shallow sea, as a narrow band of oolitic carbonate in a north-south trending barrier bar system along the eastern portion of present day Miami-Dade and Broward counties.[9] Falling sea levels eventually exposed the formation to air and rain, and rainwater percolating through the deposits replaced aragonite with calcite ($CaCO3$) and formed an indurated rock.[10] Presently, the Miami Limestone consists of two separate units—the oölitic facies (upper unit) and the bryozoan facies (lower unit).[11] The oölitic facies consists of white to orangish gray, poorly to moderately indurated, sandy limestone (grainstone) with scattered concentrations of fossils. The bryozoan facies consist of white to orangish gray, poorly to well indurated, sandy, fossiliferous limestone (grainstone and packstone).[12] The underlying strata is Fort Thompson Formation (Pleistocene) and is comprised of alternating freshwater and marine marls and limestones.[13] The Fort Thompson formation in the Miami area attains a maximum thickness of 25 m and constitutes the major part of the Biscayne aquifer.[14]

The Sangamonian Stage, which was the last interglacial period, is equivalent to Marine Isotope Stage 5e (MIS 5e), therefore placing the maximum age of the proposed impact crater at ~80 ka to ~130 ka.[15] Marine isotope stages are interchanging palaeotemperature maxima and minima, inferred from oxygen isotope data reflecting changes in the planet's temperature derived from data obtained through deep sea core samples.[16] In 1965, moreover, researchers used uranium-series dating and confirmed the age of Miami Limestone at ~130 ka.[17]

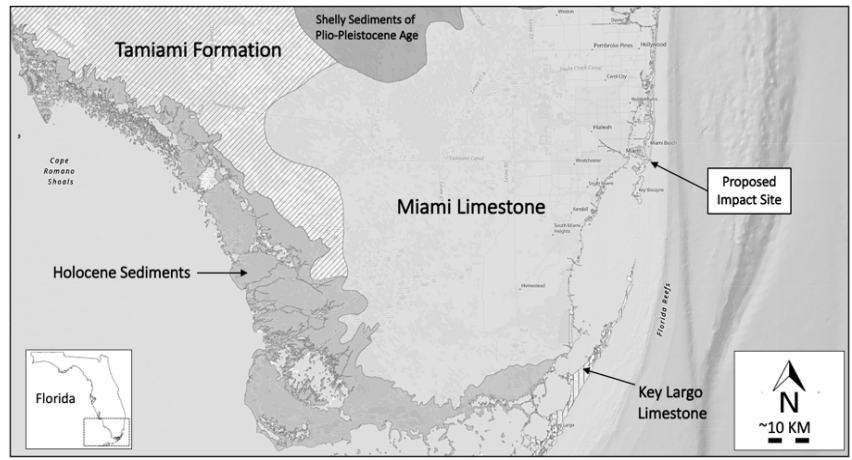

Figure 2: Geological map of South Florida (Source: ArcGIS, USGS, and Planetary Sciences, Inc.)

## AREA OF INVESTIGATION BATHYMETRY

An examination of NOAA bathymetry data (Report H11898) revealed an underwater structure illustrating morphology consistent with an impact crater produced by a hypervelocity event of extraterrestrial origin (Figure 3). The proposed impact crater has a diameter of ~650 m, has a circumference of ~2.04 km, and occupies a surface area of ~0.33 km$^2$. The prominent features, which appear more consistent with a complex crater, include a central peak and at least six outward-radiating curved ridges. The debris field ~350 m to the northwest, according to earlier research, is a proposed ejecta field associated with the impact hypothesis.[1]

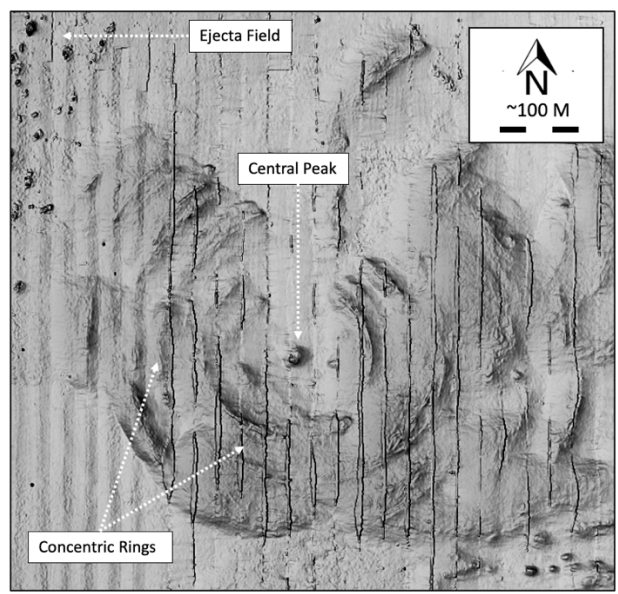

Figure 3: Bathymetry data of proposed impact site (Source: NOAA)

# IN-SITU UNDERWATER EXPEDITION

On 27 June 2020, a team of SCUBA divers surveyed the proposed impact crater. The purpose of the underwater survey was to investigate, photograph, and collect geological samples at 30 locations spread throughout the structure, which included the central peak, the northeast, southwest, northwest rings, and the proposed ejecta field (Figure 4). The underwater surveys were specifically planned for high tide at or near solar noon. Conducting the survey at or near solar noon (when the sun is directly overhead) allowed the surface area of the proposed impact site to be lit up by the sun as much as possible.

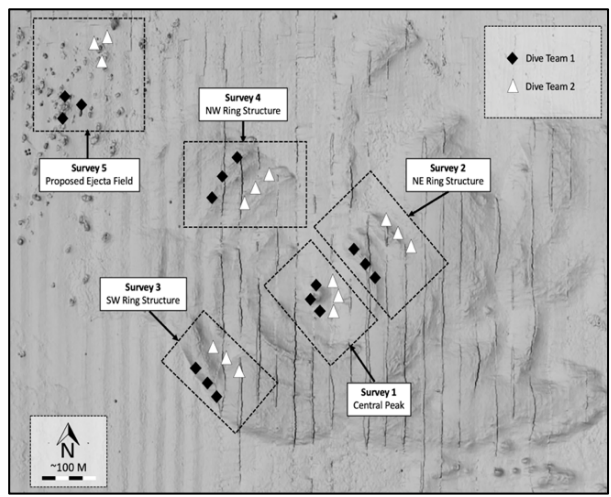

Figure 4: Underwater survey and dive plan (prepared by Planetary Sciences, Inc.)

Through the use of a computerized depth gauge, the dive team logged the proposed central peak at a depth of 8.2 m, the northeast ring at 5.79 m, the southwest ring at 6 m, and the northwest ring at 5.65 m. The recorded depths, therefore, imply that the underwater structure is a bowl-shaped depression. Furthermore, to establish whether planar formations, shatter cones, and shock metamorphic and/or other meteoritic properties were present, the dive team collected a total of 30 geological samples for analysis (Figure 5).

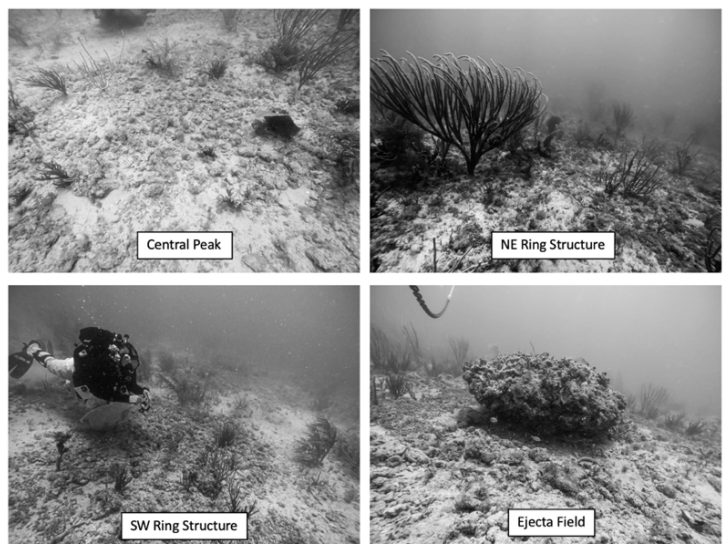

Figure 5: Underwater survey of central peak, NE and SW ring structure, and ejecta field

## ANALYSIS & INTERPRETATION

This investigation considered all possibilities for the formation of the proposed impact crater. The four competing hypotheses that could explain how the structure was formed included a controlled maritime explosion, radial lava flow from a volcano, a hypervelocity impact event of extraterrestrial origin, or a bioherm, doline, or karst formation. After investigating and analyzing all competing hypotheses, we interpret the proposed impact crater as solutional doline originating from the overlying Miami Limestone.

### Hypothesis 1: Controlled Maritime Explosion

Although various simulations of wave and debris associated with underwater explosions provided a strong argument to discredit an impact hypothesis, a central peak is not characteristic of a controlled underwater explosion.[18] Additionally, indices checks of NOAA, US Army Corps of Engineers (Jacksonville District), and US Coast Guard records returned no information regarding a controlled maritime explosion in the vicinity of 25° 44' 55.95" N and 80° 07' 12.95" W. Moreover, an examination of local historical archives dating back to at least 1903, when the dredging of Government Cut commenced, likewise rule out a controlled maritime explosion as the source for the formation of the proposed impact structure.

### Hypothesis 2: Radial Lava Flow

The USGS National Map and Volcano Hazards Program confirmed there are no known active, inactive, or ancient volcanos in the vicinity of 25° 44' 55.95" N and 80° 07' 12.95" W.[19] The Miami Limestone and Fort Thompson Formation are young geological formations and entirely non-volcanic. The geology to support a young volcano in the area, consequently, is not there. While it is probable that igneous rocks formed during the early phases of geological activity in the past (e.g., Precambrian) they are deeply buried under kilometers of sediment.[20]

### Hypothesis 3: Impact Event

The is no physical evidence to support the assertion that the proposed impact crater is the result of an impact event of extraterrestrial origin (i.e., a meteor). This confirmation is based on data gathered and analyzed during our investigation, such as the geomorphology of impact cratering on terrestrial bodies, the history of water at the impact site during the Sangamonian Stage, archival data from NASA, NOAA, and USGS, and physical evidence surveyed, recovered, and analyzed from the proposed impact site.

The formation of complex craters differs from bowl-shaped craters. Complex craters have uplifted centers, such as the proposed impact crater, but they also develop shallow floors with terraced walls. The diameters of complex craters where central peaks form, moreover, typically form in craters greater than ~3-5 km in diameter or larger.[21] Furthermore, complex-crater morphology on terrestrial planets appears to follow a consistent sequence with increasing size: small complex craters with a central peak; intermediate-sized peak ring craters, in which the central peak is replaced by a ring of peaks; and the largest craters, which encompass multiple concentric rings, known as multi-ringed basins.[22] This sequence of features is a sequence of increasing size of the crater, that of the meteorite and speed of impact (Figure 6, B-E). Geologically, therefore, there are no known small central peak craters with multiple rings—as is the case with the proposed impact crater. Second, the Miami Limestone was deposited during the interglacial Sangamon Stage, when the impact site was ~7 m above present sea level.[23] When an small impactor hits water, such as where the Miami Crater rests, the debris (ejecta) thrown out from the impact creates a unique pattern resembling a splash or mudflow—forming what is known as a rampart crater (Figure 6, F).[24] The outer edge of the debris, which usually displays lobes, is upraised. This feature gives the name "rampart" to this type of crater. While many rampart craters exist on Mars, the Nördlinger Ries impact structure in Germany is the only confirmed rampart crater on Earth.[25]

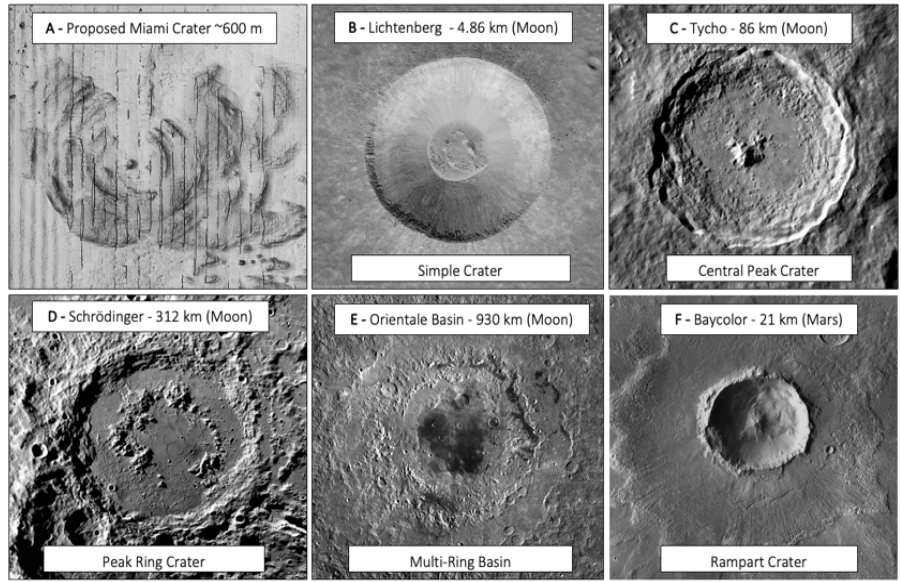

Figure 6: A comparison of the proposed impact crater (Image A) and the relation between crater sizes and complexity (Images B-F). Source: NOAA (Image A) and NASA (Images B-F)
**Not to Scale**

Furthermore, earlier researchers inferred that the "outflow of sediments running east-north-east" from the crater could be associated with the impact.[1] Bathymetric data and nautical charts from NOAA, however, indicate that the proposed impact crater rests on a designated spoilage area (dumping ground) 1.20 km south of Government Cut (Figure 7). We argue, therefore, that these sediments are not associated with an impact event, but rather material that was transported and redeposited from the dredging of Government Cut.

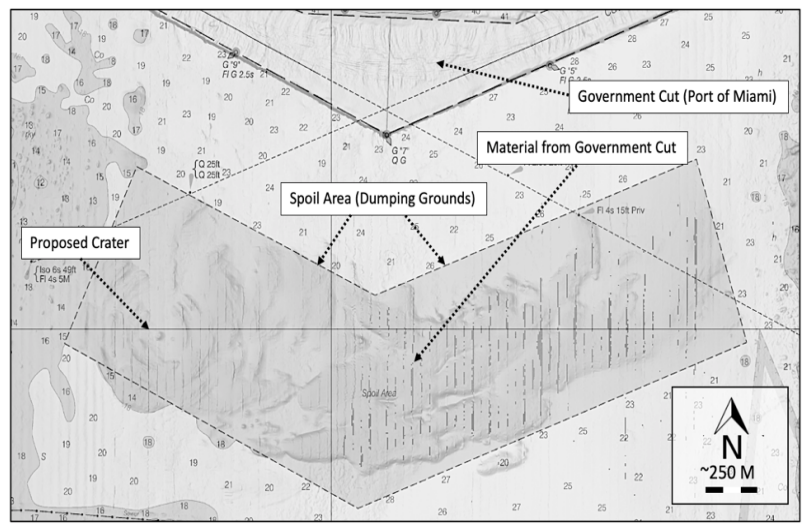

Figure 7: Spoilage area (dumping ground) for Government Cut (Source: NOAA)

## Hypothesis 4: Solutional Doline of the Underlying Limestone

An examination of comparable basinal-shaped depressions, as well as analysis of geological samples recovered *in situ*, identified the proposed impact crater as a doline formed by the uneven dissolution of the overlaying Miami Limestone. Macroscopic inspection of the 30 geological samples collected *in situ* (Figure 8 A-D) identified the samples as white to orangish gray karst limestone composed mainly of ooids, quartz sand, calcite, macroalgae and small fossils of *Modulus m.*, *Vermicularia sp.*, *Polychaete*, and *Cerithium litteratum* (Figure 9 A and 9B).[9] Most of the aragonitic ooids have been replaced by calcite and with depth these have become increasingly embedded in a matrix of crystalline calcite.[26] During examination of the samples, we found no evidence of planar formation, shatter cone, or shock metamorphic features in the sampling. Moreover, no meteoritic specimens were found at the proposed ejecta field.

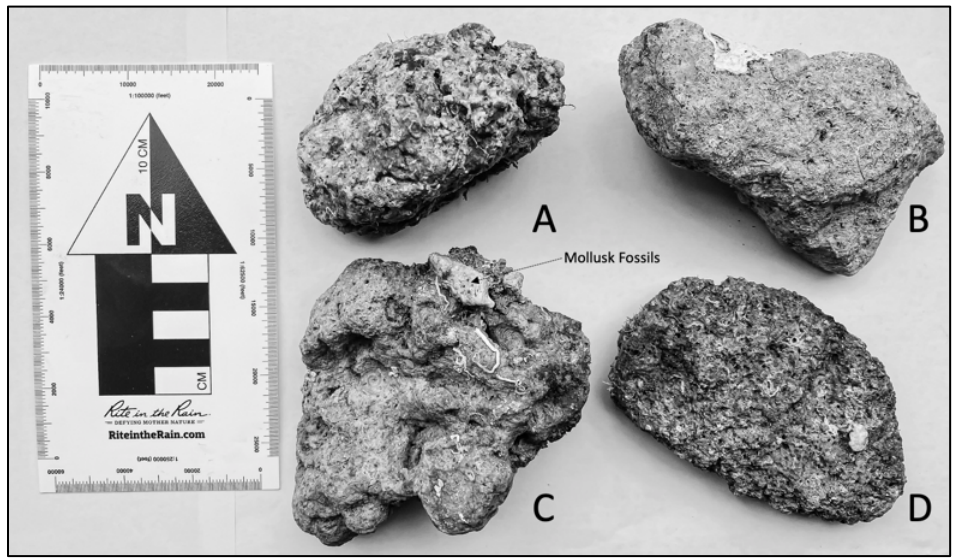

Figure 8: Samples collected at the proposed impact site.
(A) central peak, (B) NE ring, (C) NW ring, and (D) ejecta field (Source: Planetary Sciences, Inc.)

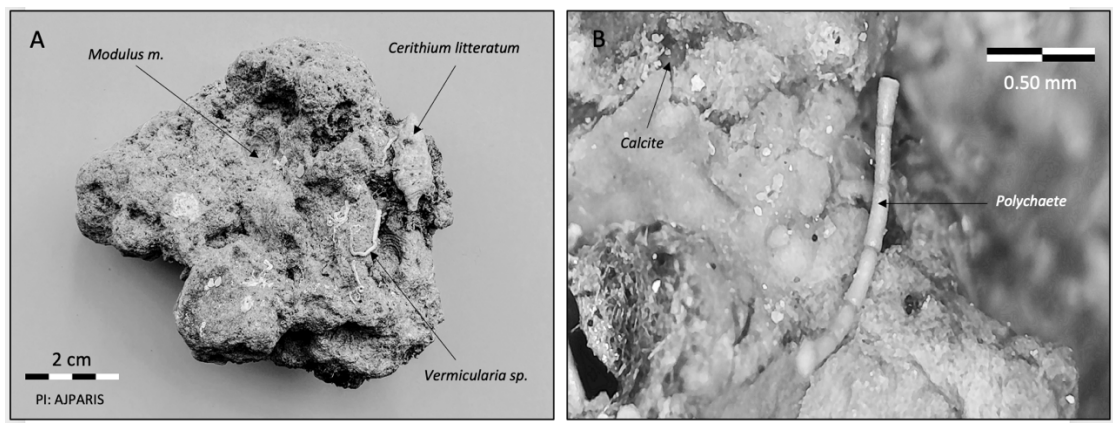

Figure 9: Macroscopic (A) and microscopic (B) images on sample collected at NW ring of proposed impact site.

Furthermore, karstification is a long-term and continuous dissolution process of water acting over carbonate rocks such as limestone and, after time, developing geological structures such as dolines.[27] Solutional dolines are known to produce broad, saucer-like depressions, particularly in a geological setting where the overlying strata is limestone.[28] Unlike a *collapsed* sinkhole (or doline), which is formed by gravitational collapse due to an underlying cavity (i.e., cave), solutional dolines can form multiple inward-dipping depressions with diameters larger than *collapsed* sinkholes or dolines. These multiple inward-dipping depressions, when observed from a position of elevation, appear as concentric outward-radiating rings analogous to complex crater morphology (Figure 10a and 10b).[29]

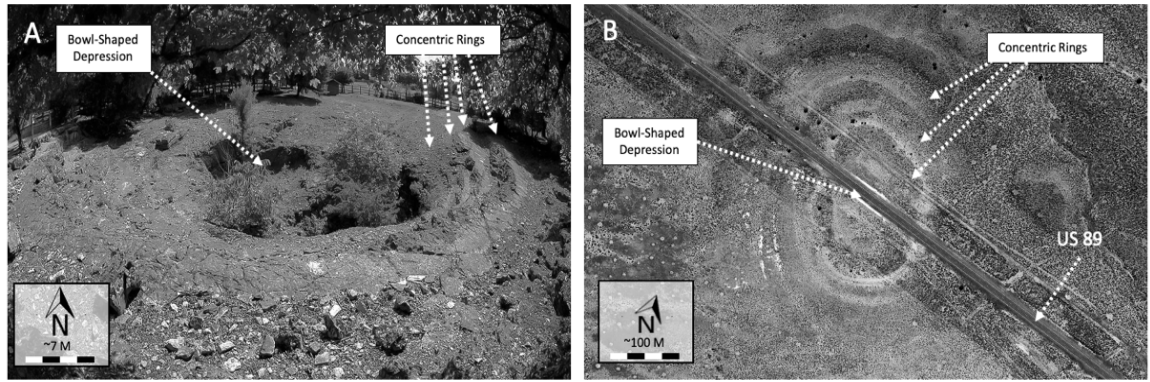

Figure 10: Analogous doline depressions with underlying limestone similar to the proposed Miami impact crater.
(A) Le Parc, France and (B) Kanab, Utah

Moreover, in a geological setting where the underlying strata is primarily limestone (i.e., Florida), dolines and sinkholes naturally form contiguously to each other. For illustration, further scrutiny of NOAA bathymetric data identified an additional doline ~400 m northeast of the proposed impact crater (Figure 11). This second doline exhibits morphology analogous to the proposed impact crater, which includes multiple concentric, outward-radiating rings.

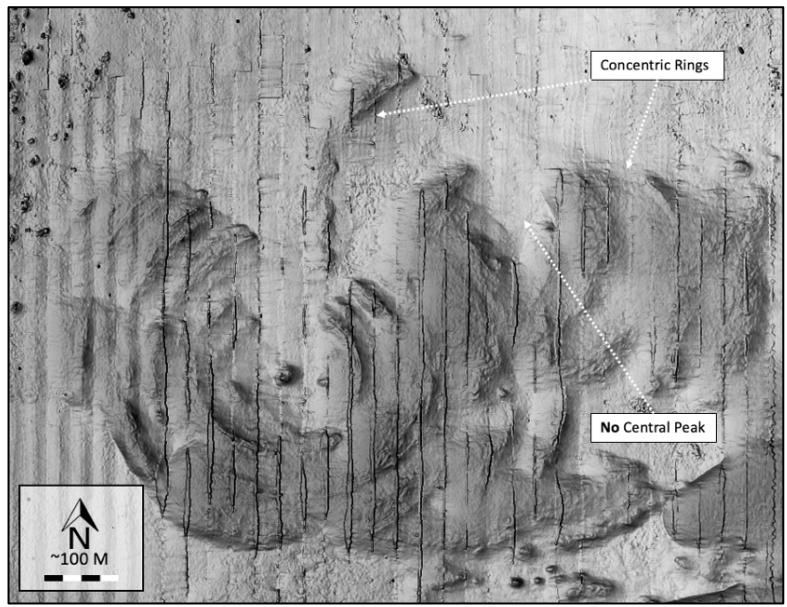

Figure 11: Similar doline with concentric rings northeast of proposed impact crater (Source: NOAA)

## CONCLUSIONS

An analysis of bathymetry data, analogous topographic depressions with concentric rings with overlying limestone, and geological samples collected *in situ* has identified the proposed Miami impact crater as a solutional doline of the overlying Miami Limestone. Additionally, other competing hypotheses for the formation of the structure, such as a controlled maritime explosion or radial lava flow from volcano, were also ruled out during our investigation. Accordingly, we do not recommend that the structure be indexed in the Earth Impact Database. Furthermore, we advise Google to update its maps and remove the current designation—*The Miami Meteorite Crater*.

## BIOGRAPHY

Antonio Paris, the Principal Investigator (PI) for this study, is the Chief Scientist at Planetary Sciences, Inc., a former Assistant Professor of Astronomy and Astrophysics at St. Petersburg College, FL, and a graduate of the NASA Mars Education Program at the Mars Space Flight Center, Arizona State University. He is the author of *Mars: Your Personal 3D Journey to the Red Planet*. His latest peer-reviewed publication is "Prospective Lava Tubes at Hellas Planitia"—an investigation into leveraging lava tubes on Mars to provide crewed missions protection from cosmic radiation. Prof. Paris is a professional member of the Washington Academy of Sciences, a member of the American Astronomical Society, and a trained SCUBA Instructor and Divemaster with the Professional Association of Divers International.

## FIELD RESEARCH CONTRIBUTERS

Ryan Robertson is the Manager of Commercial Space at Space Florida. Working in conjunction with NASA's Kennedy Space Center and the US Air Force, Ryan manages a variety of facilities spanning the State of Florida for aerospace companies to research, develop, and launch business ventures. He is currently a graduate student at American Public University studying planetary science, is a certified SCUBA diver, and has previously assisted Planetary Sciences, Inc. with research focused on Solar System bodies.

Skye Schwartz is currently an undergraduate student studying Biology at Arizona State University and the Operations Manager/Educator at Space Trek. Accepted into the NASA Solar System Ambassador program back in 2016, Skye has hosted events at schools, conferences, and NASA centers all around the United States.